\documentclass[prd,preprint,superscriptaddress,showpacs,byrevtex]{revtex4}
\newcommand{\be}{\begin{equation}}
\newcommand{\ee}{\end{equation}}
\newcommand{\bea}{\begin{eqnarray}}
\newcommand{\eea}{\end{eqnarray}}
\usepackage[dvips]{graphicx}
\begin{document}
%
%
%
\title{Infrared Behaviour of The Gluon Propagator in  
Non-Equilibrium Situations}
\author{Fred Cooper} \email{fcooper@lanl.gov}
\affiliation{T-8, Theoretical Division, Los Alamos National Laboratory,
Los Alamos, NM 87545, USA}
\author{Chung-Wen Kao} \email{kao@a35.ph.man.ac.uk}
\affiliation{Theoretical Physics Group, Department of Physics and Astronomy, University of Manchester,
Manchester,M139PL, UK}
\author{Gouranga C. Nayak} \email{nayak@shakti.lanl.gov}
\affiliation{T-8, Theoretical Division, Los Alamos National Laboratory,
Los Alamos, NM 87545, USA }

\date{\today}
\begin{abstract} 
The infrared behaviour of the medium modified gluon propagator in non-equilibrium situations
is studied in the covariant gauge  using  the Schwinger-Keldysh closed-time 
path formalism. 
It is shown that the magnetic screening mass is non-zero
at the one loop level whenever the initial gluon
distribution function is non isotropic with the assumption that the 
distribution function of the gluon is not divergent at zero transverse
momentum.
For isotropic gluon distribution functions, such as those describing local 
equilibrium,
the magnetic mass at one loop level
is zero which is consistent with finite temperature field theory results. 
Assuming that a reasonable
initial gluon distribution function can be obtained from a 
perturbative QCD calculation of minijets,
we determine these out of equilibrium  values for the initial 
magnetic and Debye screening masses
at energy densities appropriate to RHIC and LHC. 
We also compare the magnetic masses obtained here with those obtained
using finite temperature lattice QCD methods at similar temperatures
at RHIC and LHC.

\end{abstract}  
\maketitle
Experiments at RHIC (Au-Au collisions at $\sqrt s$ = 200 GeV) and
LHC (Pb-Pb collisions at $\sqrt s$ = 5.5 TeV) will
provide an excellent opportunity to produce a quark-gluon plasma in the
laboratory. There is no doubt that an energy density larger than
$\sim$ 5 GeV/fm$^3$ \cite{hijing}
will be created during these collisions but it is 
not at all clear that the partons produced following the collision will reach 
equilibrium.
The study of the equilibration of  the quark-gluon plasma is very crucial because
it determines the time evolution of all global quantities such as energy 
density, number density etc.. This study also plays a crucial role in
determining many of the potential signatures for quark-gluon plasma detection at RHIC.
The space-time evolution of  the parton gas for this non-equilibrium situation can be 
be modeled by solving semi classical  relativistic transport equations
\cite{nayak,mul,geiger,wang12,wong,gyulassy,keijo,bhal,naya}. 
Central to solving
the transport equations is what goes into the scattering kernels. Perturbative vacuum
expressions for gluon scattering suffer from severe infrared problems. One loop medium 
effects in equilibrium provide an electric (Debye) screening mass, but not a magnetic 
screening mass \cite{art}.  Thus one cannot use a one loop resummed finite temperature gluon propagator
as an approximation to the scattering Kernel because of severe infrared problems in the limit that
$p_0 =0 $ and  $ |{\vec p}|\rightarrow 0$.   
 To obtain magnetic screening in  equilibrium situations requires a non-perturbative lattice
QCD calculation. What we would like to point out here is that if we use the CTP formalism 
\cite{CTP}
with an {\it{arbitrary}} non thermal initial Gaussian density matrix then it 
is possible to obtain
at one loop a magnetic screening mass as long as the initial gluon single 
particle distribution function
$f(k_x,k_y,k_z,t_0)$ is not isotropic with the assumption that the gluon
distribution function is not divergent at zero transverse momentum. 
That is we assume that at $t=t_0$ one can write a Fourier 
decomposition of the Gluon field in terms of creation and annihilation operators. By a Bogoliubov transformation at $t=t_0$ one can always set the pair distribution functions $\langle a^\dag_\lambda ({\vec k}, t=t_0) a^\dag ({\vec q},t=t_0)_\lambda \rangle = 
\langle a_\lambda ({\vec k}, t=t_0) a ({\vec q},t=t_0)_\lambda \rangle = 0$.  Thus the propagator will have the usual vacuum part and a term which depends on the initial expectation value of the number density
\begin{equation}
 \sum_{\lambda=1}^2\langle a^\dag_\lambda ({\vec k}, t=t_0) a ({\vec q},t=t_0)_\lambda \rangle =
 f({\vec k},t_0) (2 \pi)^3 \delta({\vec k}-{\vec q}),
\end{equation}
and we have summed over the physical transverse polarizations. 
For $f$ to correspond to a physically realizable quantity the number
density as well as energy density has to be finite. Thus $f(\vec k,t)$
has to go to zero as $k \rightarrow \infty$ fast enough so that
one obtains finite number density and energy density.
For  Gaussian initial value problems, one only needs to know the two-point function at $t=t_0$.
In our following analysis, we will also need to make a quasi-adiabatic 
approximation so that we will assume, for the purpose of determining the 
initial screening masses, that the system  is time-translation invariant. 
Thus the only difference we will assume in 
our Green's functions from the usual thermal ones will be the choice of an anisotropic $f({\vec k},t_0)$ which
will replace the usual Bose Einstein distribution function. This approximation has been discussed in 
detail by Thoma and others \cite{thoma}. In our calculations,  
we will use a simple ansatz for  $f({\vec k},t_0)$ in which the parameters will be chosen to agree
with known distributions for minijet production at RHIC and LHC.  Here we are not suggesting that this
effect replaces the nonperturbative magnetic screening mass, but that the effect we are considering is of 
the same order of magnitude and already cures the infrared problems of the transport theory. 

In what follows we will examine the infrared
behaviour of the medium modified gluon propagator at one loop using the CTP formalism. 
The purpose of this paper is to study the static limit of the
longitudinal and transverse self energy of the gluon (Debye and magnetic screening
masses) and, in particular, to determine, at the one loop level, how the magnetic screening mass
depends on the initial $f({\vec k},t_0)$. Although technically the magnetic screening mass is defined
as the position of the  zero of  the inverse  propagator (i.e. in the limit $p_0=0, |{\vec p}| \rightarrow m_{sc}, m^2_{sc} = \Pi(m_{sc}^2) $) \cite{ref1},  at one 
loop the limit of the inverse propagator as  $p_0=0, |{\vec p}| \rightarrow 0$ {\em is} gauge invariant
(independent of $\xi$ for general covariant $\xi$ gauges),  and moreover this limit is the one important
for controlling the infrared properties of the collision kernel in the transport theory.  Thus in
this paper we will use the second limiting process to define the screening masses.
 At arbitrary momentum
the polarization is not in general gauge invariant at one loop. To have a gauge invariant approximation at one loop one can  make a hard momentum loop approximation as discussed in \cite{thoma,cw}.

 In particular we are interested in nonisotropic nonthermal forms for $f({\vec k},t_0)$ consistent with known minijet production results. 
Let us consider an expanding system of partons in 1+1 dimensions.  
For this purpose we introduce the flow velocity of the medium
\be
u^{\mu}~=~(\cosh\eta,0,0,\sinh\eta),
\label{flow}
\ee
where $\eta=\frac{1}{2}\ln{\frac{t+z}{t-z}}$
is the space-time rapidity and $u_{\mu}u^{\mu}~=~1$. We define
the four symmetric tensors \cite{pr,wel,land}:
\begin{eqnarray}
T_{\mu\nu}(p)&=&g_{\mu\nu}-\frac{(u\cdot p)(u_{\mu}p_{\nu}
+u_{\nu}p_{\mu})-p_{\mu}p_{\nu}-p^{2}u_{\mu}u_{\nu}}
{(u\cdot p)^{2}-p^{2}}, \nonumber \\
L_{\mu\nu}(p)&=&\frac{-p^{2}}{(u\cdot p)^{2}-p^{2}}
\left(u_{\mu}-\frac{(u\cdot p)p_{\mu}}{p^{2}}\right)
\left(u_{\nu}-\frac{(u\cdot p)p_{\nu}}{p^{2}}\right), \nonumber \\
C_{\mu\nu}(p)&=&\frac{1}{\sqrt{2[(u\cdot p)^{2}-p^{2}]}}\left[\left(u_{\mu}
-\frac{(u\cdot p)p_{\mu}}{p^{2}}\right)p_{\nu}+\left(u_{\nu}
-\frac{(u\cdot p)p_{\nu}}{p^{2}}\right)p_{\mu}\right], \nonumber \\
D_{\mu\nu}(p)&=&\frac{p_{\mu}p_{\nu}}{p^{2}}.
\label{tmunu}
\end{eqnarray}

Here $T^{\mu\nu}$ is transverse with respect to the flow-velocity but
$L^{\mu\nu}$ and $D^{\mu\nu}$ are mixtures of space-like and time-like
components. These tensors satisfy the following transversality properties
with respect to $p^{\mu}$:
\be
p_{\mu}T^{\mu\nu}(p)~=~p_{\mu}L^{\mu\nu}(p)~=~0,~~~~~~~~~~~~
p_{\mu}p_{\nu}C^{\mu\nu}(p)~=~0.
\ee

In terms of this tensor basis the gluon propagator in the covariant
gauge is given by:

\begin{equation}
\tilde{G}_{\mu\nu}(p)=-iT_{\mu\nu}(p)\tilde{G}^{T}(p)
-iL_{\mu\nu}(p)\tilde{G}^{L}(p)
-i\xi D_{\mu\nu}(p)\tilde{G}^{D}(p),
\end{equation}
where $\tilde{G}^T$, $\tilde{G}^L$, $\tilde{G}^D$ correspond to $T$, $L$ and
$D$ components respectively of the full gluon propagator at the one loop level. The last part $\tilde{G}^D(p)$ is identical
to the vacuum part \cite{pr} and hence we do not consider it any more.
There are separate
Dyson-Schwinger equations for the different components of the CTP matrix Green's functions that do not couple with
each other. 
These equations can be written in the form 
\begin{equation}
\left[\tilde{G}^{T,L}(p)\right]_{ij}=\left[G^{T,L}(p)\right]_{ij}
+\sum_{l,k}\left[G^{T,L}(p)\right]_{il}
\cdot\left[\Pi^{T,L}(p)\right]_{lk}\cdot \left[\tilde{G}^{T,L}(p)\right]_{kj}.
\end{equation}
Here $i,j,k,l~=~+,-$ are the CTP contour labels, and suppression of Lorentz and color
indices in the above equation is understood.

In the Keldysh rotated representation of the CTP formalism, in terms of retarded, advanced and symmetric
Green's functions we have instead
\begin{equation}
\tilde{G}^{T,L}_{R,A}(p)=G^{T,L}_{R,A}(p)+G^{T,L}_{R,A}(p)
\cdot\Pi^{T,L}_{R,A}(p)\cdot \tilde{G}^{T,L}_{R,A}(p).
\end{equation}

The straightforward solution of the above equation is given by:
\begin{equation}
\tilde{G}^{T,L}_{R,A}(p)=\frac{G^{T,L}_{R,A}(p)}{1-G^{T,L}_{R,A}(p)\cdot
\Pi^{T,L}_{R,A}(p)}=\frac{1}{p^2-\Pi^{T,L}_{R,A}(p) \pm i sgn(p_{0})\epsilon},
\end{equation}
where the self-energy contains the medium effects.
Similar but more complicated equations are obtained for the resummed
symmetric Green's functions
\begin{equation}
\begin{array}[c]{c}
\tilde{G}^{T,L}_{S}(p)=\left[1+2f(\vec{p})\right]sgn(p_{0})[\tilde{G}^{T,L}_{R}(p)
-\tilde{G}^{T,L}_{A}(p)]\\
+\left(\Pi_{S}^{T,L}(p)-(1+2f(\vec{p}))sgn(p_{0})[\Pi^{T,L}_{R}(p)-\Pi^{T,L}_{A}(p)]\right)
\times \tilde{G}^{T,L}_{R}(p)\times \tilde{G}^{T,L}_{A}(p).
\end{array}
\end{equation}

For the purposes of obtaining the correct kernel for the Boltzmann equation, we need the medium
improved  Feynman
propagator for the gluon at one loop level which is just one component $\{++\}$  of the matrix 
Green's function of the CTP formalism. 

$G_F(p)  \equiv \left[\tilde{G}(p)\right]_{++}$, can be written as:

\be
\left[\tilde{G}(p)\right]_{++}=\frac{1}{2}
\left[\tilde{G}_{S}(p)+\tilde{G}_{A}(p)+\tilde{G}_{R}(p)\right], 
\ee
where 
where $G_A, G_R, G_S$ stands for advanced, retarded and symmetric green's function respectively.
In the above equation the '+' sign refers to the upper branch in the closed-time path.
Using the relations of the various self energies one finds
\cite{thoma,cw}:
\be
\tilde{G}_{++}(p)=\frac{p^2-Re \Pi_R(p)+\frac{1}{2}Im \Pi_S(p)}{
(p^2-Re \Pi_R(p))^2+(Im \Pi_R(p))^2},
\label{resum}
\ee
where $Re \Pi_R(p)$ and $Im \Pi_R(p)$ are the 
real and imaginary part of the retarded
self energy. These self-energies have both longitudinal and transverse parts 
which, in the static limit ($p_0=0, |\vec p | \rightarrow 0$), give Debye and
magnetic screening masses of the gluon respectively. In the above equation
$\Pi_S(p)$ is the the symmetric part of the self energy.

\begin{figure}
   \centering
   \includegraphics{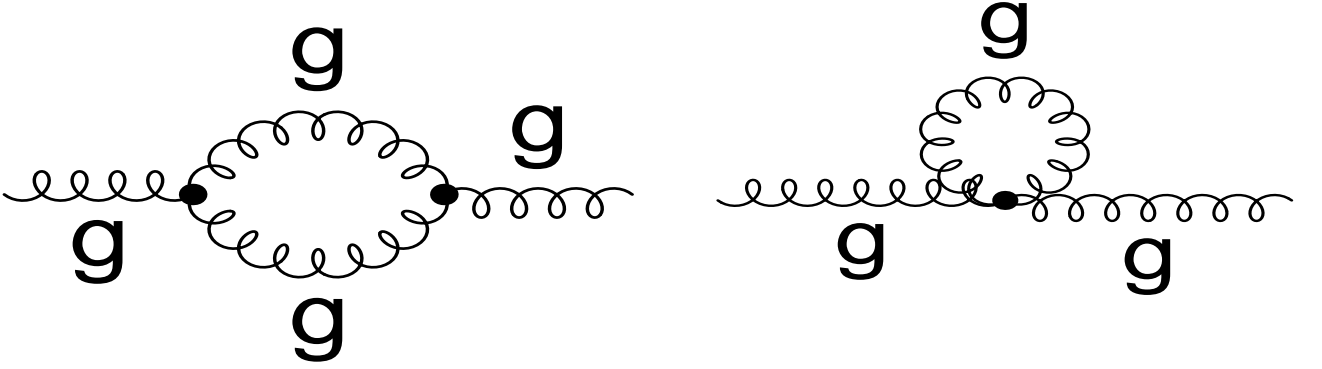}
   \caption{One Loop Graphs for the Gluon Self-Energy}
\label{fig:graphs}
\end{figure}

To obtain the infrared behaviour of this propagator we need to find
the static limit of the gluon self energy for an anisotropic $f({\vec k} ,t)$ corresponding
to the initial distribution function expected from the parton model. 
In a frozen ghost formalism \cite{land,petro}, the gluon self
energy is obtained from the gluon loop and tadpole loop as shown
in Fig. \ref{fig:graphs}. The ghost does not contribute to the medium effect in this
formalism because the initial density of states are chosen to be that
of the physical gluons. All the effects of the ghost are present in the vacuum. 

The general expressions for the real and imaginary part of the 
gluon self energy in non-equilibrium
in covariant gauge for an expanding gluonic medium have been derived in a 
previous paper \cite{cw}.
Here we examine the static limit of these self-energies which play 
crucial
roles to obtain a finite collision integral to study equilibration
of the quark-gluon plasma at RHIC and LHC. The general expressions
for the real part of 
the longitudinal and transverse self energy
of the gluon loop are given by:

\bea
Re\Pi^{L}_{Gl;R}(p)=&&\frac{g^{2}}{2}\delta_{ab}N_{c}
\int\frac{d^{3}q}{(2\pi)^{3}}~[\frac{1}{2|\vec q|}
[[f(\vec q)G(q,p)]_{q^0=|\vec q|} +[f(- \vec q)G(q,p)]_{q^0=-|\vec q|} ] 
~~~~~~~~~~~~~~~~~~~~~~~~~\nonumber \\
&&+\frac{1}{2|\vec p -\vec q|} 
[[f(\vec p - \vec q)G(p-q,p)]_{p^0-q^0=|\vec p - \vec q|}
+[f(-\vec p + \vec q)G(p-q,p)]_{p^0-q^0=-|\vec p - \vec q|}]]
\label{pil}
\eea
and
\bea
Re\Pi^{T}_{Gl;R}(p)=&&\frac{g^{2}}{2}\delta_{ab}N_{c}
\int\frac{d^{3}q}{(2\pi)^{3}}~[\frac{1}{2|\vec q|}
[[f(\vec q)H(q,p)]_{q^0=|\vec q|} +[f(- \vec q) H(q,p)]_{q^0=-|\vec q|} ] 
~~~~~~~~~~~~~~~~~~~~~~~~~\nonumber \\
&&+\frac{1}{2|\vec p -\vec q|} 
[[f(\vec p - \vec q)H(p-q,p)]_{p^0-q^0=|\vec p - \vec q|}
+[f(- \vec p + \vec q)H(p-q,p)]_{p^0-q^0=-|\vec p - \vec q|}]]
\label{pit}
\eea
where
\begin{eqnarray}
G(q,p)&=&\frac{1}{(p^0-q^0)^2-(\vec p - \vec q)^2}~
[\frac{8p^{2}}{(u\cdot p)^{2}-p^{2}}[((u\cdot q)
-\frac{(u\cdot p)(q\cdot p)}{p^2})^2] \nonumber \\
&-&\left[(p+q)^{2}\right] 
\left[\frac{2(q\cdot p)(p\cdot u)^3}{(q \cdot u)p^2((p \cdot u)^2-p^2)}
-\frac{(q\cdot p)^2(p\cdot u)^2}{(q \cdot u)^2p^2((p \cdot u)^2-p^2)}
-\frac{(p\cdot u)^2}{((p \cdot u)^2-p^2)}
\right] \nonumber \\
&-&4p^{2}+8\frac{(q\cdot p)(u\cdot p)}{(u\cdot q)}-4
\frac{(q\cdot p)^{2}}{(u\cdot q)^{2}} \nonumber \\
&+& (\xi-1)\frac{(p \cdot u)^2 ( (q \cdot p)^2-2 (q \cdot u) 
(p \cdot u) (q \cdot p)+ (q \cdot u)^2p^2)}{(q \cdot u)^2
(p^2- (p \cdot u)^2)}].
\end{eqnarray}
and
\begin{eqnarray}
H(q,p)&=& 
\frac{1}{(p^0-q^0)^2-(\vec p - \vec q)^2}~
 [8\frac{(u\cdot q)(q\cdot p)(u\cdot p)}{(u\cdot p)^{2}-p^2}
-4\frac{(q\cdot p)^{2}}{(u\cdot p)^{2}-p^2} 
-4\frac{p^2(q\cdot u)^{2}}{(u\cdot p)^{2}-p^2} \nonumber \\
&-&\left[(p+q)^{2}\right]\left[1-
\frac{(q\cdot p)(u\cdot p)}{(u\cdot q)((u\cdot p)^{2}-p^2)}
+\frac{(q\cdot p)^{2}}{2(u\cdot q)^{2}((u\cdot p)^{2}-p^2)}+
\frac{p^2}{2((u\cdot p)^{2}-p^2)}\right] \nonumber \\
&-&4p^{2}+8\frac{(q\cdot p)(u\cdot p)}{(u\cdot q)}-4
\frac{(q\cdot p)^{2}}{(u\cdot q)^{2}} \nonumber \\
&+&\frac{(\xi-1)(p^4(- (q \cdot p)^2+2 (q \cdot u) (p \cdot u) (q \cdot p) 
+(q \cdot u)^2 (p^2-2 (p \cdot u)^2)))}
{2(q \cdot u)^2( (q \cdot p)-p^2)(p^2- (p \cdot u)^2)}].
\end{eqnarray}

The general expression for the tadpole loop contribution is given by:
\begin{eqnarray}
&&\Pi^{L}_{Ta;R}(p)={g^{2}}\delta_{ab}N_{c}
\int\frac{d^{3}q}{(2\pi)^{3}}\frac{f(\vec{q})}{2|\vec{q}|} 
 \nonumber \\
&&
[3-2\frac{(u\cdot p)(q\cdot p)}{p^{2}(u\cdot q)}+\frac{p^{2}}{(u\cdot p)^{2}-p^{2}}(1-\frac{(u\cdot p)(q\cdot p)}{p^{2}(u\cdot q)})^{2}]|_{q_{0}=|\vec{q}|}
\nonumber \\
&+&\frac{f(-\vec{q})}{2|\vec{q}|} [3-2\frac{(u\cdot p)(q\cdot p)}{p^{2}(u\cdot q)}+\frac{p^{2}}{(u\cdot p)^{2}-p^{2}}(1-\frac{(u\cdot p)(q\cdot p)}{p^{2}(u\cdot q)})^{2}]|_{q_{0}=-|\vec{q}|}
 \nonumber \\
\label{tdl}
\end{eqnarray}
and

\begin{eqnarray}
&&\Pi^{T}_{Ta;R}(p)=g^{2}\delta_{ab}N_{c}
\int\frac{d^{3}q}{(2\pi)^{3}}\frac{f(\vec{q})}{2|\vec{q}|} 
[1
+\frac{(u\cdot p)(q\cdot p)}{(u\cdot q)((u\cdot p)^{2}-p^{2})}  \nonumber \\
&& -\frac{(q\cdot p)^{2}}
{2(u\cdot q)^{2}((u\cdot p)^{2}-p^{2})}
-
\frac{p^{2}}{2((u\cdot p)^{2}-p^{2})}
]|_{q_{0}=|\vec{q}|}
\nonumber \\
&&+\frac{f(-\vec{q})}{2|\vec{q}|}[1
+\frac{(u\cdot p)(q\cdot p)}{(u\cdot q)((u\cdot p)^{2}-p^{2})}-\frac{(q\cdot p)^{2}}
{2(u\cdot q)^{2}((u\cdot p)^{2}-p^{2})}-\frac{p^{2}}{2((u\cdot p)^{2}-p^{2})}
]|_{q_{0}=-|\vec{q}|}. \nonumber \\
\label{tdt}
\end{eqnarray}
To simplify these equations in the infrared limit
we expand $f(\vec q -\vec p)$ as:
$f(\vec q -\vec p) = f(\vec q) - 
\vec p \cdot \vec{\nabla_q} f(\vec q)$
and neglect the higher order gradients. Similarly we expand
$|\vec p - \vec q|$ as: 
$|\vec p - \vec q|= |\vec q| (1 - \frac{\vec p \cdot \hat q}{|\vec q|})$.
In the static limit (first taking $p_0=0$ then using 
$|\vec p | \rightarrow 0$), and in the rest frame ($u_0=1, \vec u =0$)
we obtain from Eq. (\ref{pil}): 
\begin{equation}
Re\Pi^{L}_{Gl;R}(p_{0}=0,|\vec{p}|\rightarrow 0)
=-2g^{2}\delta_{ab}N_{c}
\left[\int\frac{d^{3}q}{(2\pi)^{3}}(\frac{\hat{p}\cdot \nabla_{q}f(\vec{q})}{\hat{p}\cdot \hat{q}})
+\int\frac{d^{3}q}{(2\pi)^{3}}\frac{f(\vec{q})}{|\vec{q}|}\right]
\end{equation}
and from Eq. (\ref{tdl}):
\begin{equation}
\Pi^{L}_{Ta;R}(p_{0}=0)=2{g^{2}}\delta_{ab}N_{c}
\int\frac{d^{3}q}{(2\pi)^{3}}\frac{f(\vec{q})}{|\vec{q}|}. 
\end{equation}
Adding both the above
equations we get the expression for the Debye screening mass:
\bea
m_D^2=[Re\Pi^{L}_{Gl;R}(p_{0}=0,|\vec{p}|\rightarrow 0)]
+[Re\Pi^{L}_{Ta;R}(p_{0}=0,|\vec{p}|\rightarrow 0)] \nonumber \\
=-6g^2
\int\frac{d^{3}q}{(2\pi)^{3}}(\frac{\hat{p}\cdot \nabla_{q}f(\vec{q})}{\hat{p}\cdot \hat{q}})
\label{md}
\eea
which is the real part of the longitudinal self energy. $N_C=3$ is used.
This equation was obtained by various authors \cite{biro}.
Similarly in the static limit and in the rest frame we get from Eq.
(\ref{pit}):
\begin{equation}
Re\Pi^{T}_{Gl;R}(p_{0}=0,|\vec{p}|\rightarrow 0)
=g^{2}\delta_{ab}N_{c}
\int\frac{d^{3}q}{(2\pi)^{3}}[\frac{f(\vec{q})}{|\vec{q}|}
\cdot[\frac{3}{2}(\hat{q}\cdot \hat{p})^{2} -\frac{1}{2}]
+\frac{\hat{p}\cdot \nabla_{q}f(\vec{q})}{\hat{p}\cdot \hat{q}}
\cdot[1-(\hat{q}\cdot \hat{p})^{2}]]
\end{equation}
and from Eq. (\ref{tdt}):
\begin{equation}
\Pi^{T}_{Ta;R}(p_{0}=0)=g^{2}\delta_{ab}N_{c}
\int\frac{d^{3}q}{(2\pi)^{3}}\frac{f(\vec{q})}{|\vec{q}|}\cdot[\frac{3}{2}-\frac{1}{2}(\hat{q}\cdot \hat{p})^{2}].
\end{equation}

Adding both the above
equations we get the expression for the magnetic screening mass:
\bea
m_g^2= [Re\Pi^{T}_{Gl,R}(p_{0}=0,|\vec{p}|\rightarrow 0)]
+ [Re\Pi^{T}_{Ta,R}(p_{0}=0,|\vec{p}|\rightarrow 0)] \nonumber \\
=3g^{2}
\int\frac{d^{3}q}{(2\pi)^{3}}[\frac{f(\vec{q})}{|\vec{q}|}\cdot [1+(\hat{q}\cdot \hat{p})^{2}]
+\frac{\hat{p}\cdot \nabla_{q}f(\vec{q})}{\hat{p}\cdot \hat{q}}
\cdot[1-(\hat{q}\cdot \hat{p})^{2}]]
\label{mg}
\eea
which is the real part of the transverse self energy. For the 
imaginary part of the gluon self energy at one loop we get: 
$Im\Pi_{GL:R}^{T,L}=0$, in the static limit. Note that the above 
formula uses the medium part of the self energy containing 
a gluon loop and a tadpole loop which appears in the 
resummed gluon propagator (Eq. (\ref{resum})). 
The expression for the Debye screening mass, we obtained 
(see Eq. (\ref{md})), is the same as that obtained
by various authors \cite{biro}. 
The expression we obtain here for the magnetic 
mass for a non-equilibrium gluon
distribtution function is new (see Eq. (\ref{mg})). 
There is no approximation 
present in the derivation of Eqs. (\ref{md}) and (\ref{mg}). The static limit 
results are gauge invariant.

For an isotropic gluon distribution function 
$f(|\vec{q}|)$ we get from the eq. (\ref{md})
\begin{equation}
m_D^2
=\frac{6 g^2}{\pi^2} \int dq q f(q)
\label{md2}
\end{equation}
where $q=|\vec q|$ and from eq. (\ref{mg})
\be
m_g^2=0.
\label{mg2}
\ee
Furthermore for the special isotropic case when 
the system is described by an equilibrium 
Bose-Einstein distribution function for the gluon
Eqs. (\ref{md}) and (\ref{mg}) give: 
\bea
m_D^2=g^2T^2,~~~~~~~~~~~~~{\rm and}~~~~~~~~~~~~~~~m_g^2=0
\label{mdg}
\eea
respectively. These  results (Eq. (\ref{mdg})) are identical to those
 obtained by using finite temperature field
theory in QCD assuming that the system is in thermal equilibrium 
\cite{art,wel,bz}.
It is interesting to note that the magnetic mass is not only zero at one
loop level in equilibrium (Eq. (\ref{mdg})) but it is also zero for any
isotropic non-equilibrium gluon distribution function (see Eq. (\ref{mg2})). 
Only when the distribution function is non-isotropic one
gets a non-zero contribution to the magnetic screening mass with the assumption
that the distribution function of the gluon
is not divergent at zero transverse momentum (see below).  
This result is not particular to QCD but also will 
be true for QED when the distribution function is non-isotropic. This
is explicitly calculated in \cite{fred2} for the QED case
where we have shown that
we exactly get the same formula for the magnetic screening mass in QED 
as we obtained in this paper for QCD (gluon loop)
except that $N_c g^2 \propto e^2$. 

Before considering the situation at RHIC and LHC let us consider an
example where there is momentum anisotropy in the transverse and longitudinal
momentum distribution. 
For this purpose we work in the cylindrical coordinate
system: $(q_t,\phi,q_z)$. From  Eq. (\ref{md}) we get: 
\bea
{m_D^2} =-\frac{6g^2}{(2\pi)^3}
\int d^{2}q_t \int dq_z |\vec q| 
(\frac{\hat{p}\cdot \nabla_{q}f(\vec{q})}{ \hat{p} \cdot \vec{q}}). 
\label{mdi}
\eea
~From this equation we realize that when $f(\vec{q})$ is isotropic, 
the dependence
on $\hat{p}$ drops out and we obtain eq. (\ref{md2}).  For anistropic $f$ the
mass depends on the direction of $\hat p$. In what follows we will assume 
$\hat p$ is along the
transverse direction and give values only for this direction. 
Similar results can be obtained for
the longitudinal choice.    
Assuming $\hat p$ is along  the transverse direction we find 

\bea
m^2_{Dt}=-\frac{6g^2}{(2\pi)^3}
\int dq_t \int d \phi \int dq_z  \sqrt{q_t^2 + q_z^2}
\frac {\partial f(q_t, \phi, q_z)}{\partial q_t}.
\eea
Integrating by parts in $q_t$ we get:
\bea
m^2_{Dt}=
 \frac{3 g^2}{4\pi^3} [\int dq_t q_t \int d\phi \int \frac{dq_z}{|\vec q|}
~f(q_t, \phi, q_z)
+\int d \phi \int d q_z ~
[|q_z| f(q_t, q_z, \phi)]_{q_t=0}].
\label{mdi2}
\eea
For an equilibrium distribution function of the form $f_{eq} ~= ~
\frac{1}{e^{\sqrt{q_x^2+q_y^2+q_z^2}/T}-1}~ 
=~\frac{1}{e^{\sqrt{q_t^2+q_z^2}/T}-1}$ we get from the above
equation:
\be
m^2_{Dt}= \frac{g^2T^2}{2}+\frac{g^2T^2}{2}~=~g^2T^2
\ee
which is the correct result obtained by using finite temperature QCD.

Similarly, 
changing to $q_t, \phi, q_z$ coordinate system we get from Eq. (\ref{mg})
\bea
m_g^2 =\frac{3g^{2}}{8\pi^3}
\int dq_t q_t \int d\phi \int dq_z [\frac{f(\vec{q})}{|\vec{q}|}\cdot [1+
\frac{(\vec{q}\cdot \hat{p})^{2}}{|\vec q|^2}]
+|\vec q| \frac{\hat{p}\cdot \nabla_{q}f(\vec{q})}{\hat{p}\cdot \vec{q}}
\cdot[1-\frac{(\vec{q}\cdot \hat{p})^{2}}{|\vec q|^2}]].
\label{mgi}
\eea
When $\hat p$ points in the transverse direction and we again perform
partial integration over $q_t$ to obtain:
\bea
{m_g}_t^2 =\frac{3g^{2}}{8\pi^3}[
2 \int dq_t q_t \int d\phi 
\cos^2\phi \int dq_z \frac{f(q_t,q_z, \phi)}{|\vec{q}|}~
-~\int d \phi \int d q_z ~[|q_z| f(q_t,q_z,\phi)]_{q_t=0}].
\label{mgi2}
\eea
For an equilibrium distribution function of the form:
$f_{eq} ~= ~ \frac{1}{e^{\sqrt{q_x^2+q_y^2+q_z^2}/T}-1}~ 
=~\frac{1}{e^{\sqrt{q_t^2+q_z^2}/T}-1}$ 
the above equation gives:
\bea
{m_g}_t^2 =\frac{3g^{2}}{(2\pi)^3}[
(4\pi) \frac{\pi^2 T^2}{6}-
(4\pi) \frac{\pi^2 T^2}{6}]=0,
\label{mgi3}
\eea
which is consistent with finite temperature QCD results.

Before proceeding to compute the initial magnetic screening mass at RHIC and
LHC situations we will adopt a non-isotropic test distribution function
to compute the magnetic screening mass from the formula given by 
eq. (\ref{mgi2}). We choose a non-isotropic test distribution function of the
form:
\be
f~=~\frac{1}{e^{\sqrt{q_t^2+hq_z^2}/T}-1}
\label{nd}
\ee
where $h$ is a parameter for non-isotropy. For $h~=~1$ we get the usual
Bose-Einstein distribution function. Using the above non-isotropic 
distribution function we plot the magnetic screening mass from the 
eq. (\ref{mgi2}) in Fig. 2. It can be seen from Fig. 2 that for $h~=~1$ 
(corresponding to Bose-Einstein distribution function) we get
$m_{gt}~=~0$ and for $h~\neq ~ 1$ (corresponding to non-isotropic
distribution function) we get a non-zero magnetic screening mass. 

\begin{figure}
   \centering
   \includegraphics{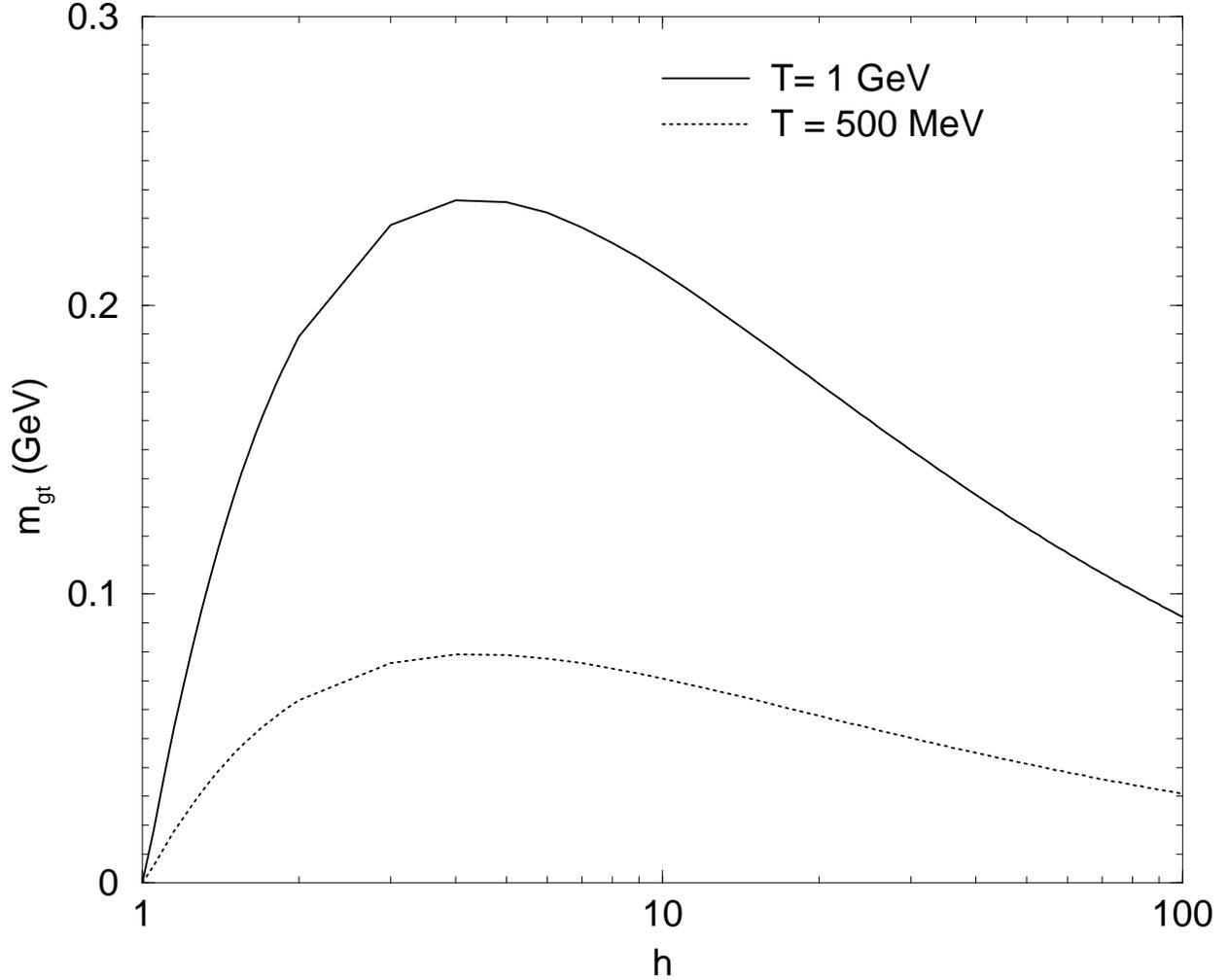}
   \caption{The transverse component of the magnetic screening mass as
obtained from eq. (\ref{mgi2}) by using 
a non-isotropic distribution function of the form: 
$f~=~\frac{1}{e^{\sqrt{q_t^2+hq_z^2}/T}-1}$ as a function of the
non-isotropy parameter $h$. Note that for $h~=~1$ the distribution
function becomes Bose-Einstein and hence the magnetic screening mass 
is found to be zero from eq. (\ref{mgi2}).}
\label{phs}
\end{figure}

Now we consider the realistic situation at RHIC and LHC.
For the situation at RHIC and LHC where the parton distribution function
at $t=t_0$, $f(\vec q, t_0)$, describes
an out of equilibrium situation we can compute the value of these
screening masses assuming the distribution function can be described 
by the parton model result for minijets. 
We note that to compute the second term in rhs of eqs. (\ref{mdi2}) and
(\ref{mgi2}) we need to know the form of $[f(q_t,q_z)]_{q_t=0}$ 
and its behaviour at RHIC and
LHC. In this paper we are considering the minijet distribution function
which are computed by using pQCD applicable above $q_t$ = 1(2) GeV at RHIC
(LHC) which are obtained by saturation arguments as
studied by several authors \cite{pmin}. We mention here that pQCD is not 
applicable for small $q_t$, for example below 1(2) GeV at RHIC(LHC).
If one calculates the pQCD minijet 
production the $q_t$ distribution behaves as: $\propto ~q_t^{-\alpha}$
where $\alpha \sim 4 $ for high $q_t$ and $\sim 2$ for low $q_t$.
If one applies pQCD at small $q_t$ the distribution function $f(q_t,q_z)$
is singular at $q_t=0$. However, for very low $q_t$
pQCD formulas are not applicable and hence it is not obvious that
distribution will be singular at $q_t$=0 at RHIC and LHC.  
For $q_t=0$ the only computation
available at the moment is from the McLerran-Venugopalan model \cite{alex}
where it is shown that at $q_t$=0 the gluon distribution behaves
as a constant with respect to $q_t$ and does not behave as 
$q_t^{-\alpha}$. In this cases we may assume that the gluon distribution
does not diverge at $q_t$=0 in the realistic situation at RHIC and LHC.
In such situations where $f(q_t=0,y, \phi)$ is not singular ($q_z=q_t \sinh y$
for gluon minijets) one can neglect the boundary term:
\bea
&&[\sqrt{q_t^2 + q_z^2}  f(q_t,q_z,\phi)]|_{q_t=0} ^{q_t=\infty} = 
- [|q_z| f(q_t, q_z,\phi)]_{q_t=0}= \nonumber \\
&&-  [q_t |\sinh y| f(q_t, y,\phi)]_{q_t=0}= 0
\label{bc1}
\eea
since for massless minijets
\be
q_z=q_t \sinh y =0 ~~~~~{\rm when} ~~~~~~~~q_t=0 ~~~~~{\rm for~~ finite}~~~ y.  
\label{bc2}
\ee
Here $y$ is the momentum rapidity of the minijet parton.
This boundary condition is {\em not} true in general, 
and in particular not true for a thermal 
distribution since a thermal distribution function: 
$f_{eq}=\frac{1}{e^{q_t \cosh y}/T-1}$ is divergent at $q_t$=0.
However if the gluon distribution at $q_t$ =0 behaves as a constant
at $q_t$=0 at RHIC and LHC initial situations \cite{alex} then our vanishing
boundary condition should be valid at RHIC and LHC.
In any case a non-perturbative analysis of gluon distribution
at $q_t$=0 is beyond the scope of this paper. If the gluon distribution
behaviour at $q_t$=0 is found to be divergent in any non-perturbaive 
calculation unlike the case in \cite{alex} then the values reported
in this paper might change. We have computed the Debye and magnetic screening
masses in this paper above $q_t$=1(2) GeV at RHIC(LHC) which is
similar to the calculations done by several authors for the
Debye screening mass \cite{all1} where they have adopted
similar cut-off values for the minijet momentum in their calculations.

With the above arguments and with the vanishing boundary 
conditions (eqs. (\ref{bc1}) and (\ref{bc2})) we get from eq. (\ref{mdi2}),
after
changing to the rapidity variables: $\frac{dq_z}{|\vec q|}= dy$:
\bea
{m_D^2}_t =
 \frac{3 \alpha_s}{\pi^2} \int dq_t q_t \int d\phi \int dy ~f(q_t, \phi,y),
\label{mdf1}
\eea
where $f(q_t, \phi, y)$ is the non-isotropic gluon distribution
function. For a cylindrically symmetric system we get:
\bea
{m_D^2}_t =
 \frac{6 \alpha_s}{\pi} \int dq_t q_t \int dy ~f(q_t, y).
\label{mdf}
\eea
This is exactly the same equation used by several authors \cite{nayak,all} 
in the context
of minijet plasma equilibration in heavy-ion collisions at RHIC and LHC.
Similarly using the same vanishing boundary condition (eqs. (\ref{bc1})
and (\ref{bc2})) we get for 
the magnetic screening mass from eq. (\ref{mgi2}):
\bea
{m_g^2}_t =
\frac{3 \alpha_s}{\pi} \int dq_t q_t \int dy ~f(q_t, y),
\label{mgf}
\eea
for cylindrically symmetric distribution function $f(q_t, y)$.
It can be noted that in eqs. (\ref{mdf}) and
(\ref{mgf}) one should not use equilibrium distribution function or any other
distribution function which does not obey the vanishing 
boundary condition as stated in eqs. (\ref{bc1}) and (\ref{bc2}).

For conditions pertinent to RHIC and LHC we use the minijet gluon distribution
function to evaluate the Debye and magnetic screening masses.
At high energy the minijet cross section can be calculated by
using perturbative QCD (pQCD). The leading order minijet cross section is given by:
\begin{equation}
\sigma_{jet} = \int dp_t \int dy_1 \int dy_2 {{2 \pi p_t} \over {\hat{s}}} 
\sum_{ijkl}
x_1~ f_{i/A}(x_1, p_t^2)~ x_2~ f_{j/A}(x_2, p_t^2)~
\hat{\sigma}_{ij \rightarrow kl}(\hat{s}, \hat{t}, \hat{u}).
\label{jet}
\end{equation}
Here $x_1$ and $x_2$ are the light-cone momentum fractions carried by
the partons $i$ and $j$ from the projectile and the target,
respectively, $f$ are the bound-nucleon structure functions and $y_1$
and $y_2$ are the rapidities of the scattered partons. The symbols
with carets refer to the parton-parton c.m. system. The
$\hat{\sigma}_{ij \rightarrow kl}$ are the elementary pQCD parton cross
sections. As we will be considering a gluon system we include the
dominant gluon production cross sections at the partonic level which are
given by:
\begin{equation}
\hat{\sigma}_{gq \rightarrow gq} = {{\alpha_s^2} \over
{ \hat{s}}} ({\hat{s}^2+\hat{u}^2}) [ 
{{1} \over {\hat{t}^2}}
- {{4} \over {9\hat{s}\hat{u}}}], 
\nonumber
\end{equation}
and
\begin{equation}
\hat{\sigma}_{gg \rightarrow gg} = {{9 \alpha_s^2} \over
{2 \hat{s}}} [ 3 - {{\hat{u}\hat{t}} \over {\hat{s}^2}}
 - {{\hat{u}\hat{s}} \over {\hat{t}^2}}
 - {{\hat{s}\hat{t}} \over {\hat{u}^2}}].
\nonumber
\end{equation}
Here $\alpha_s$ is the strong coupling constant and
\begin{equation}
\hat{s}=x_1 x_2 s = 4 p_t^2 ~{\rm cosh}^2
\left ( {{y_1-y_2} \over {2}} \right ).
\nonumber 
\end{equation}
The rapidities $y_1$, $y_2$ and the momentum fractions $x_1$, $x_2$ are
related by,
\begin{equation}
x_1=p_t~(e^{y_1}+e^{y_2})/{\sqrt{s}}, \hspace{0.5cm}
x_2=p_t~(e^{-y_1}+e^{-y_2})/{\sqrt{s}}.
\nonumber
\end{equation}
The limits of integrations are given by:
\bea
p_{min} \leq p_t \leq \frac{\sqrt{s}}{2 \cosh y_1},~~~~~~ 
-ln({\sqrt{s}/{p_t}-e^{-y_1}}) \leq y_2 \leq
ln({\sqrt{s}/{p_t}-e^{y_1}}),
\eea
 with
\be
\vert {y_1} \vert \leq ln(\sqrt{s}/{2 p_{min}} + \sqrt{s/{4 p_{min}^2}
-1}). 
\ee
In the above equations $p_{min}$ is the minimum transverse momentum
above which minijet production is computed by using pQCD.
We multiply the above minijet cross
sections by  a $K$ factor $K=2$  to account for the higher order
$O(\alpha_s^3)$ contributions. The minimum transverse momentum
above which the minijets are computed via pQCD are of the order
of $p_{min} \sim$ 1 GeV at RHIC and $\sim$ 2 GeV at LHC \cite{pmin}. 
These values are energy dependent and are 
obtained from the saturation arguments.
We take $p_{min}$ = 1 GeV at RHIC and
2 GeV at LHC for our computations.
The minijet cross section (Eq.\ (\ref{jet}))
can be related to the total number of partons $(N)$ by
\begin{equation}
N^{jet}=T(0) ~\sigma_{jet},
\label{number}
\end{equation}
where $T(0)= 9A^2/{8\pi R_A^2}$ is the total number of
nucleon-nucleon collisions per unit area for central collisions
\cite{kaja}. Here $R_A=1.1 A^{1/3}$ is the nuclear radius. 
A rough estimate of the initial volume in which these initial
partons are formed at RHIC and LHC can be given by:
$V_0 = \pi R_A^2 \tau_0$, where the partons are
assumed to be spread by a length $\tau_0=1/p_{min}$.
Assuming that the partons are uniformly distributed in the coordinate
space (but non-isotropic in momentum space)
we can easily extract a phase-space gluon distribution function
of the gluon 
from the total number of gluon minijets from Eq. (\ref{number}).
The initial distribution function of the gluon is then given by: 
\be
f(p_t,y_1)=\frac{1}{\pi R_A^2 \tau_0} dN^{jet}/d^3p
\label{f0}
\ee
where
\begin{equation}
d^3p=d^2p_t dp_z=p_t ~d^2p_t ~{\rm cosh} {y_1} ~{dy_1}.
\label{d3p}
\end{equation}

Using the above minijet initial gluon distribution function 
in Eq. (\ref{mdf}) and Eq. (\ref{mgf}) we get:

\bea
&&{m_D^2}_t= \nonumber \\
&& \frac{T(0)}{\pi^2 R_A^2 \tau_0}
 6 K \alpha_s \int dp_t  \int dy_1 
\int dy_2 {{1} \over {\hat{s} \cosh y_1}} 
\sum_{ijkl}
x_1 f_{i/A}(x_1, p_t^2) x_2~f_{j/A}(x_2, p_t^2)
\hat{\sigma}_{ij \rightarrow kl}(\hat{s}, \hat{t}, \hat{u})~~,
\label{mdff}
\nonumber \\
\eea
for the Debye screening mass and:
\bea
&&{m_g^2}_t = \nonumber \\
 &&\frac{T(0)}{\pi^2 R_A^2 \tau_0}
 3 K \alpha_s \int dp_t \int dy_1 
\int dy_2 {{1} \over {\hat{s} \cosh y_1}} 
\sum_{ijkl}
x_1 f_{i/A}(x_1, p_t^2) x_2 f_{j/A}(x_2, p_t^2)
\hat{\sigma}_{ij \rightarrow kl}(\hat{s}, \hat{t}, \hat{u})~~,
\nonumber \\
\label{mgff}
\eea
for the magnetic screening mass of the gluon at the one loop level.
Note that in the above equation $\alpha_s$ occurs outside the $p_t$
integration and hence a scale has to be defined, at which this
coupling constant has to be determined.
For this purpose we take $\alpha_s$ as $\alpha_s(<p_t^2>)$
where the momentum scale $<p_t^2>$ is defined by:
\bea
<p_t^2>=\frac{1}{\sigma^{jet}}
\int dp_t p_t^2 \int dy_1 \int dy_2 {{2 \pi p_t} \over {\hat{s}}} 
\sum_{ijkl}
x_1~ f_{i/A}(x_1, p_t^2)~ x_2~ f_{j/A}(x_2, p_t^2)~
\hat{\sigma}_{ij \rightarrow kl}(\hat{s}, \hat{t}, \hat{u}),
\nonumber \\
\label{pt2}
\eea
where $\sigma^{jet}$ is defined by the Eq. (\ref{jet}).

\begin{figure}
   \centering
   \includegraphics{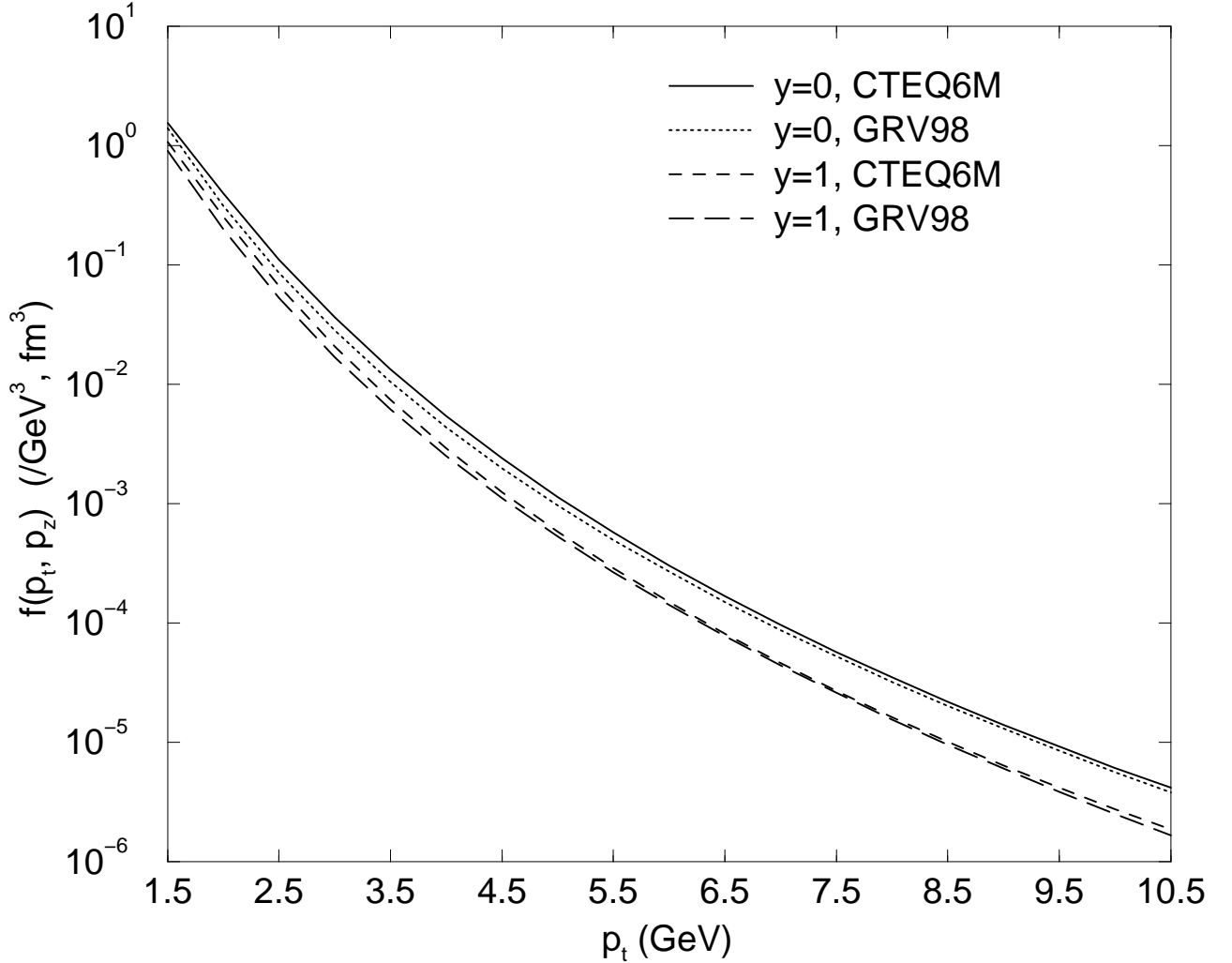}
   \caption{Initial gluon distribution function at RHIC energies as a 
function of $p_t$}
   \label{fig:ptrhic}
\end{figure}

In this paper we will be using both GRV98 \cite{grv98}
and CTEQ6M \cite{cteq6} parametrizations for the gluon and quark
structure functions inside free proton with EKS98 \cite{eks98} 
parametrizations for the nuclear modifications. 
In Fig. \ref{fig:ptrhic} we present
the results of the initial gluon distribution function (see Eq. (\ref{f0}))
at RHIC
as a function of the transverse momentum of the gluon for different
values of the rapidities. The rapidity $y$ is related to the 
longitudinal momentum $p_z$ via: $p_z=p_t \sinh y$.

\begin{figure}
   \centering
   \includegraphics{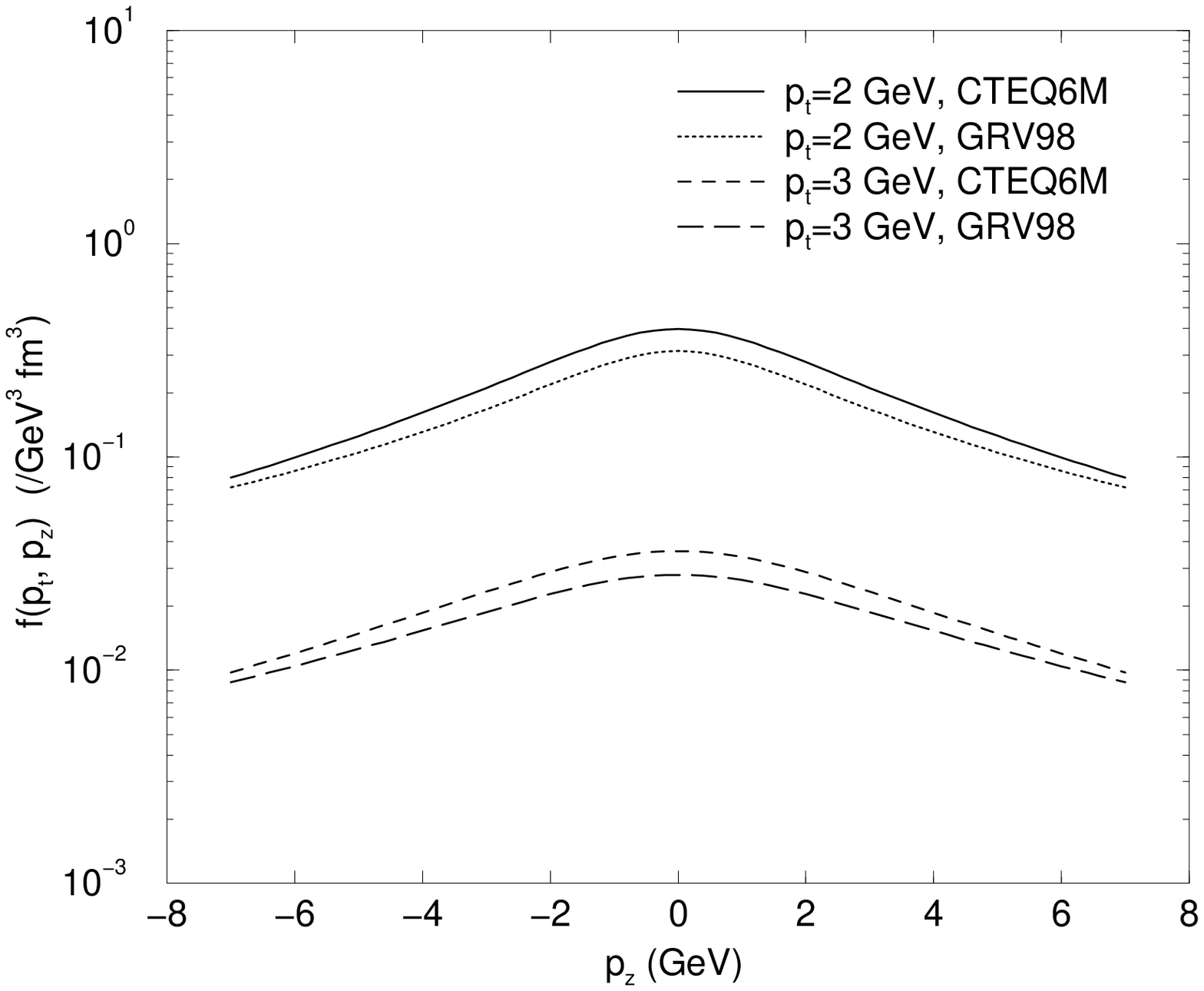}
   \caption{Initial gluon distribution function at RHIC energies as a 
function of $p_z$}
   \label{fig:pzrhic}
\end{figure}

 We present the longitudinal
momentum distribution of the initial gluon minijet distribution function
at RHIC in Fig. \ref{fig:pzrhic} for different values of $p_t$.

Using these gluon minijet distribution functions in eqs. (\ref{mdff}) and
(\ref{mgff}) we get for RHIC:
\be
{m_{D}}_t= 116 ~{\rm MeV} ~~~~~~~~{\rm and} ~~~~~~~~~~~~~~~{m_{g}}_t= 82 ~{\rm
MeV }
\ee
and for LHC: 
\be
{m_{D}}_t= 150~{\rm MeV} ~~~~~~~~~~~~{\rm and} ~~~~~~~~~~{m_{g}}_t= 105 ~{\rm 
MeV }
\ee
by using GRV98 structure functions along with EKS98. The coupling
constant $\alpha_s(<p_t^2>)$ is found to be 0.287 at RHIC and 0.214 at LHC.
If higher order contribution to the minijet production would have not
been taken into account then our results of screening mass would have
been $\sqrt{K(=2)}$ times less than the above values.
The above masses may be lower bounds to the actual values as we have
used a lower transverse momentum cutoff for minijets in order that pQCD to be
applicable. However, the gluon
distribution may be dominant at lower $p_t$ \cite{alex} and hence
the magnitude of the screening mass may increase
if one can include the soft partons into the gluon distribution
function. The values we reported in this paper are for
gluon minijet distribution functions at RHIC and LHC
with $p_{min}$ greater than 1 and 2 GeV respectively. 

Let us now look at the equilibrium
situation. 
Note that at one loop level we get (see Eq. (\ref{mdg})): 
$m_D^2 = g^2 T^2$ and $m_g^2=0$ in equilibrium. Since the one loop magnetic
mass is zero in equilibrium 
we will compare our results with the magnetic mass
which is obtained by using non-perturbative methods. 
The magnetic mass obtained by using non-perturbative
methods in equilibrium is given by: $m_g^2 =\frac{3}{2}(0.255 g^2 T)^2$,
see \cite{wang12,nonp}. Assuming a temperature of about 500 MeV at RHIC
and by using the coupling constant value $\alpha_s$=0.287
at RHIC (see above) we get: $m_D=gT$= 950 MeV
and $m_g =\sqrt{\frac{3}{2}}(0.255 g^2 T)$= 563 MeV. Assuming $T=1000$ MeV
at LHC and using the LHC coupling constant $\alpha_s= 0.214$ we obtain 
$m_D =1.639$ GeV and $m_g= 840$ MeV.  In obtaining these masses, one has
integrated over all momentum ranges of the equilibrium distribution functions.
For example, if one uses a Bose-Einstein distribution
function in Eq. (\ref{md2}) and then integrates  
from ($p_{min} \rightarrow \sqrt s/2$)
then we obtain $m_D=486$MeV at RHIC for 
T=500 MeV and $\alpha_s$=0.287. Similarly for
LHC one  obtains $m_D=840$ MeV for T=1 GeV and
$\alpha_s$= 0.214. 

Note that these values are of the same order as that
obtained by using the non-equilibrium
distribution functions at RHIC and LHC. Since the gluon
distribution function may be dominant at lower $p_t$ 
the magnitude of the screening mass might increase
if one can include the soft partons into the gluon distribution
function \cite{bhal,naya,others,larry1,denis,roberts,bla}. 
The values we reported in this paper are for
gluon minijet distribution functions at RHIC and LHC
with $p_{min}$ greater than 1 and 2 GeV respectively. 
Note that due to the asymmetry we have computed a specific
component ($\hat p$ in transverse direction) of the 
Debye (${m_D}_t$) and magnetic (${m_g}_t$) screening mass. 
If one computes the values in all directions their values
may be even higher.
Similar situations hold for magnetic screening masses at RHIC and LHC.
As the magnetic mass is a non-perturbative calculation at 
equilibrium and ours is a one loop calculation at non-equilibrium,
we expect that a non-perturbative non-equilibrium calculation
might give a higher magnetic screening mass. The argument is similar
to the study of non-perturbative calculation for Debye screening
mass at finite temperature \cite{kajn}.

To summarize, we have applied the closed-time path 
formalism to non-equilibrium situations in QCD expected at RHIC and
LHC energies to study the infrared behaviour of the one loop gluon
self energy. We have followed a frozen ghost formalism 
where the initial density of states consists of physical
gluons and the ghost is only present in the vacuum level.
In the infrared limit of the gluon
self energy we obtain a non-vanishing magnetic screening mass
of the gluon at one loop level for non-isotropic gluon distribution
functions with the assumption that 
the distribution function of the gluon is not divergent at zero transverse 
momentum. At RHIC and LHC we assumed that the gluon distribution is 
not divergent at $q_t$=0 which is supported by the computation done in
\cite{alex}. With this approximation
we then applied pQCD above $q_t$ =1(2) GeV at RHIC(LHC)
and obtain a reasonable initial non-equilibrium
gluon-minijet distribution function.
Using this non-isotropic gluon minijet distribution function above
$q_t$ = 1(2) GeV at RHIC(LHC) we predicted the values of the
magnetic and Debye screening masses at the initial time.

\acknowledgements
We thank Tanmoy Bhattacharya, Larry McLerran, Emil Mottola, E. V. Shuryak
and Raju Venugopalan for useful discussions.


\newpage

\begin{thebibliography}{aaaaaaaaaaaaaaa}

\bibitem{hijing} X. N. Wang, Phys. Rep. 280 (1997) 287;
X. N. Wang and Miklos Gyulassy, Phys. Rev. D 44
(1991) 1991; K. J. Eskola and K. Kajantie, Z. Phys. C 75 (1997) 515;
A. Krasnitz and R. Venugopalan, 
Phys. Rev. Lett. 84 (2000) 4309; K. Geiger, Phys. Rep. 258 (1995) 237;
K. Geiger and B. Muller, Nucl. Phys. B 369 (1992) 600;
 N. Hammon, H. Stoecker, W. Greiner, Phys. Rev C 61 (2000) 014901.
\bibitem{nayak} G. C. Nayak, A. Dumitru, L. McLerran and W. Greiner, 
Nucl. Phys. A 687 (2001) 457.
\bibitem{mul} R. Baier, A. H. Mueller, D. Schiff and D. T. Son,
Phys. Lett. B 502 (2001) 51, A. H. Mueller, Phys. Lett. B 475 (2000) 220.
\bibitem{geiger} K. Geiger, Phys. Rep. 280 (1995) 237;
K. Geiger and J. I. Kapusta, Phys. Rev. D47, (1993) 4905.
\bibitem{wang12}
H. Heiselberg, X.N. Wang, Nucl. Phys. B 462:389-414 (1996). 
\bibitem{wong}
S.H.M. Wong, Phys. Rev C54 (1996) 2588; Phys. Rev. C56 (1997) 1075.
\bibitem{gyulassy}
M. Gyulassy, Y. Pang, B. Zhang, Nucl. Phys. A 626 (1997) 999;
B. Zhang, Comput. Phys. Commun. 109 (1998) 193.
\bibitem{keijo} A. Hosoya and K. Kajantie, Nucl. Phys. B 250 (1985) 666.
\bibitem{bhal}
R.S. Bhalerao, G.C. Nayak, Phys. Rev. C 61 (2000) 054907.
\bibitem{naya} G. C. Nayak and V. Ravishankar, Phys. Rev D55 (1997) 6877;
Phys. Rev. C58 (1998) 356.
\bibitem{art} R. D. Pisarski, Phys. Rev. D47 (1993) 5589; 
E. Braaten and R. D. Pisarski, Nucl. Phys. B 339 (1990) 310;
T. Altherr, Phys. Lett. B341 (1995) 325; R. Baier, M. Dirks, K. Redlich and
D. Schiff, Phys. Rev. D56 (1997) 2548; and references therein.
\bibitem{CTP}
J.~Schwinger,
   J.\ Math.\ Phys.\ \textbf{2} (1961) 407;
   P.~M.\ Bakshi and K.~T.~Mahanthappa,
   J.\ Math.\ Phys. \textbf{4} (1963) 1;
   \textbf{4} (1963) 12;
   L.~V.~Keldysh,
   Zh.\ Eksp.\ Teo.\ Fiz.\ \textbf{47} (1964) 1515;
   [Sov.\ Phys.\ JETP \textbf{20} (1965) 1018];
   G.~Zhou, Z.~Su, B.~Hao and L.~Yu,
   Phys.\ Rep.\ \textbf{118} (1985) 1;
\bibitem{ref1} A. K. Rebhan, Phys. Rev. D48 (1993) R3967.
\bibitem{thoma} for example see, 
M. H. Thoma, hep-ph/0010164, 10 th Jyvaskyla Summer School, 
Jyvaskyla, Finland, 31 Jul-18 Aug (2000);
S. Mrowczynski and M. H. Thoma, Phys. Rev. D 62 (2000) 036011.
\bibitem{pr} N. P. Landsman and Ch.G. van Weert, Phys. Rep. 145 (1987) 141.
\bibitem{wel} H. A. Weldon, Phys. Rev. D 26 (1982) 1394, and references
therein.
\bibitem{land} P. V. Landshoff and A. Rebhan, Nucl. Phys. B 383 (1992)
607 and Erratum, {\it ibid} 406 (1993) 517.
\bibitem{cw} C-W. Kao, G. C. Nayak and W. Greiner, Phys. Rev. D66 (2002)
034017, hep-ph/0102153.
\bibitem{petro} M. D'Attanasio and M. Pietroni, Nucl. Phys. B 498 (1997) 443.
\bibitem{biro} See for example: T. S. Biro, B. Muller and X.N. Wang,
Phys. Lett. B 283 (1992) 171; H. Satz and D. Srivastava, Phys. Lett.
B 475 (2000) 225; T. S. Biro,
Int. J. Mod. Phys. E1 (1992) 39; and references therein.
\bibitem{all} J. Bjoraker and R. Venugopalan, Phys. Rev. C63 (2001)
024609; J. Serreau and D. Schiff, JHEP 0111 (2001) 039; 
R. Baier, A. H. Mueller, D. Schiff and D. T. Son,
Phys. Lett. B502 (2001) 51.
\bibitem{fred2} F. Cooper, C-W. Kao and G. C. Nayak, hep-ph/0207370.
\bibitem{mw} K. J. Eskola, B. Muller, X-N. Wang, Phys. Lett. 
B374 (1996) 20.
\bibitem{pmin} A. H. Mueller, Nucl. Phys. B 572 (2000) 227; K. J. Eskola,
K. Kajantie, P. V. Ruuskanen and K. Tuominen, Nucl. Phys. B 570 (2000)
379.
\bibitem{kaja} K. J. Eskola, K. Kajantie and J. Lindfors, Nucl. Phys. B
323 (1989) 37.
\bibitem{grv98} M. Glueck, E. Reya and A. Vogt, Euro. Phys. J. C5 (1998) 461.
\bibitem{cteq6} J. Pumplin, D. R. Stump, J. Huston, H. L. Lai,
P. Nadolsky and W. K. Tung, hep-ph/0201195.
\bibitem{eks98} K. J. Eskola, V. J. Kolhinen and P. V. Ruuskanen, Nucl. Phys.
B 535 (1998) 351; K. J. Eskola, V. J. Kolhinen and C. A. Sagado,
Euro. Phys. J. C9 (1999) 61.
\bibitem{nonp} See, 
T.S. Biro and B. Muller, Nucl. Phys. A 561 (1993) 477, and
references therein; J. Ruppert, G. C. Nayak, D. D. Dietrich, H. Stoecker
and W. Greiner, Phys. Lett. B520 (2001) 233.
\bibitem{others}
Y. Kluger, J.M. Eisenberg, B. Svetitsky, F. Cooper and E. Motolla,
Phys. Rev. Lett. 67 (1991) 2427; F. Cooper, J. M. Eisenberg, Y.
Kluger, E. Motolla and B. Svetitsky, Phys. Rev. D 48 (1993) 190;
J.M. Eisenberg Phys. Rev. D 36 (1987) 3114; {\it ibid} D 40 (1989)
456; T. S. Biro, H.B. Nielsen and J. Knoll, Nucl Phys. B 245
(1984) 449; M. Herrmann and J. Knoll, Phys. Lett. B 234 (1990)
437; D. Boyanovsky H.J. de Vega, R. Holman, D.s. Lee and A. Singh,
Phys. Rev. D 51 (1995) 4419; H. Gies Phys. Rev. D 61 (2000)
085021.
\bibitem{larry1} L. Mclerran and R. Venugopalan, Phys. Rev. D 49 (1994) 2233,
Phys. Rev. D 49 (1994) 3352; Yu. Kovchegov and A. H. Mueller, Nucl. Phys.
B 529 (1998) 451; A. Kovner, L. McLerran and H. Weigert, Phys. Rev D
52 (1998) 3809, Phys. Rev. D 52 (1998) 6231; Y. V. Kovchegov, E. Levin and L. 
McLerran, hepph/9912367.
\bibitem{denis} D. D. Dietrich, G. C. Nayak and W. Greiner, Phys. Rev. D64
(2001) 074006; hep-ph/0009178; hep-ph/0202144, 
J. Phys. G (in press); 
G.C. Nayak, D. D. Dietrich and W. Greiner, Rostock 2000/Trento
2001, Exploring Quark Matter 71-78, hep-ph/0104030.
\bibitem{roberts}
C. D. Roberts and S. M. Schmidt, Prog. Part. Nucl. Phys. 45
Suppl.1:1-103, 2000; V. Vinik, {\it et. al}, Eur. Phys. J.C22 (2001) 341; 
J.C.R. Bloch, C. D. Roberts and S. M. Schmidt, Phys. Rev. D61 (2000) 117502;
J. C. R. Bloch, {\it et. al}, Phys. Rev. D (1999) 116011.
\bibitem{bla} 
J-P. Blaizot and E. Iancu, Phys. Rept. 359 (2002) 355 ;
Nucl. Phys. B570 (2000) 326; Nucl. Phys. B417 (1994) 608.

\bibitem{bz} J-P. Blaizot, E. Iancu and R. R. Parwani, Phys. Rev. D52 (1995)
2543.

\bibitem{kajn} K. Kajantie {\it et al.}, Phys. Rev. Lett. 79 (1997) 3130.
\bibitem{alex} A. Krasnitz, Y. Nara and R. Venugopalan, Phys. Rev. Lett. 87 
(2001) 192302.
\bibitem{all1} K. J. Eskola, B. Muller and X-N. Wang, Phys. Lett. 
B374 (1996) 20 and references therein. 


\end{thebibliography}
\end{document}